\documentclass[aps,prl,twocolumn,10pt,superscriptaddress,amsmath]{revtex4-1}

\usepackage[utf8x]{inputenc}

\usepackage{amsmath}
\usepackage{amsfonts}
\usepackage{amssymb}
\usepackage{graphicx}

\usepackage[bookmarks=true,colorlinks=true,citecolor=blue,linkcolor=red,urlcolor=magenta]{hyperref}

\begin{document}

\title{Superluminal neutrinos and quantum cross-correlation theory of neutrino source location}

\author{V.D. Rusov}
\affiliation{Department of Theoretical and Experimental Nuclear Physics, Odessa National Polytechnic University, Odessa, Ukraine}
\email{siiis@te.net.ua}

\author{V.I. Vysotskii}
\affiliation{Department of Theoretical Radiophysics, Taras Shevchenko National University, Kiev, Ukraine}

\author{V.A. Tarasov}
\author{I.V. Sharph}
\author{V.P. Smolyar}
\author{T.N. Zelentsova}
\author{K.K. Merkotan}
\author{E.P. Linnik}
\author{M.E. Beglaryan}
\affiliation{Department of Theoretical and Experimental Nuclear Physics, Odessa National Polytechnic University, Odessa, Ukraine}

\begin{abstract}
Based on the developed cross-correlation theory for remote location of neutrino source with two-detector setup for neutrino detection the modification of arrangement of the OPERA experiment is suggested. Within the framework of computing experiment based on the OPERA experimental data we show that the use of this theory makes it possible not only to determine with high accuracy the delay time between neutrino signals but to eliminate the errors of blind analysis by which all necessary time corrections for determination of signal "technologically unremovable" delay time between CERN and GSL are performed.
\end{abstract}

\maketitle

\section{Introduction}
Based on data processing over the period 2009-2011 the OPERA collaboration has recently announced the results about possible evidence for superluminal propagation of neutrinos \cite{bib-1}.

As the recent developments show \cite{bib-4}, the most weak point of the OPERA data interpretation \cite{bib-1} associated with the superluminal neutrino detection, is so-called blind analysis by which all necessary time corrections are performed and thereby the total signal "technologically irremovable" delay time between CERN and GSL is determined.

In this paper we propose the cross-correlation theory for remote location of neutrino source (e.g. CERN Target chamber) based on the two-detector (e.g. GSL and New Detector GSL-clone) setup for neutrino detection, which makes it possible not only to determine the delay time between neutrino signals, but to eliminate the errors of blind analysis as well.

\section{Cross-correlation theory of neutrino source location}

The proposed method for remote location of neutrino source is based on an intensity interference correlation of neutrino signal sequences from the same source measured by two or more spaced-apart detectors.

First of all, let us say a few words about the cause of hidden but really existent interference of neutrino intensities in the OPERA-experiment. It is known, that within the framework of this experiment muon neutrinos are produced due to pion and kaon decays, which, in its turn, are produced in proton-proton collisions at a graphitic target in CERN. It is obvious that in the OPERA experiment interference correlations of identical particles \cite{bib-5,*bib-5a,*bib-5b,*bib-5c,*bib-5d}, which a priori take place in hadron-hadron interactions \footnote{It is appropriate to mention here a deep and impressive physical analogy between pair correlations of identical particles \cite{bib-5} produced in nuclear interactions and correlations of photons from optical sources, for example, in the experiments of the Hanbury Brown-Twiss type with two detectors \cite{bib-6}.}, are "genetically inherited" by muon neutrinos. Just these inherited properties of neutrino became the basis for the proposed theory.

\begin{figure}
\begin{center}
\includegraphics[width=0.6\linewidth]{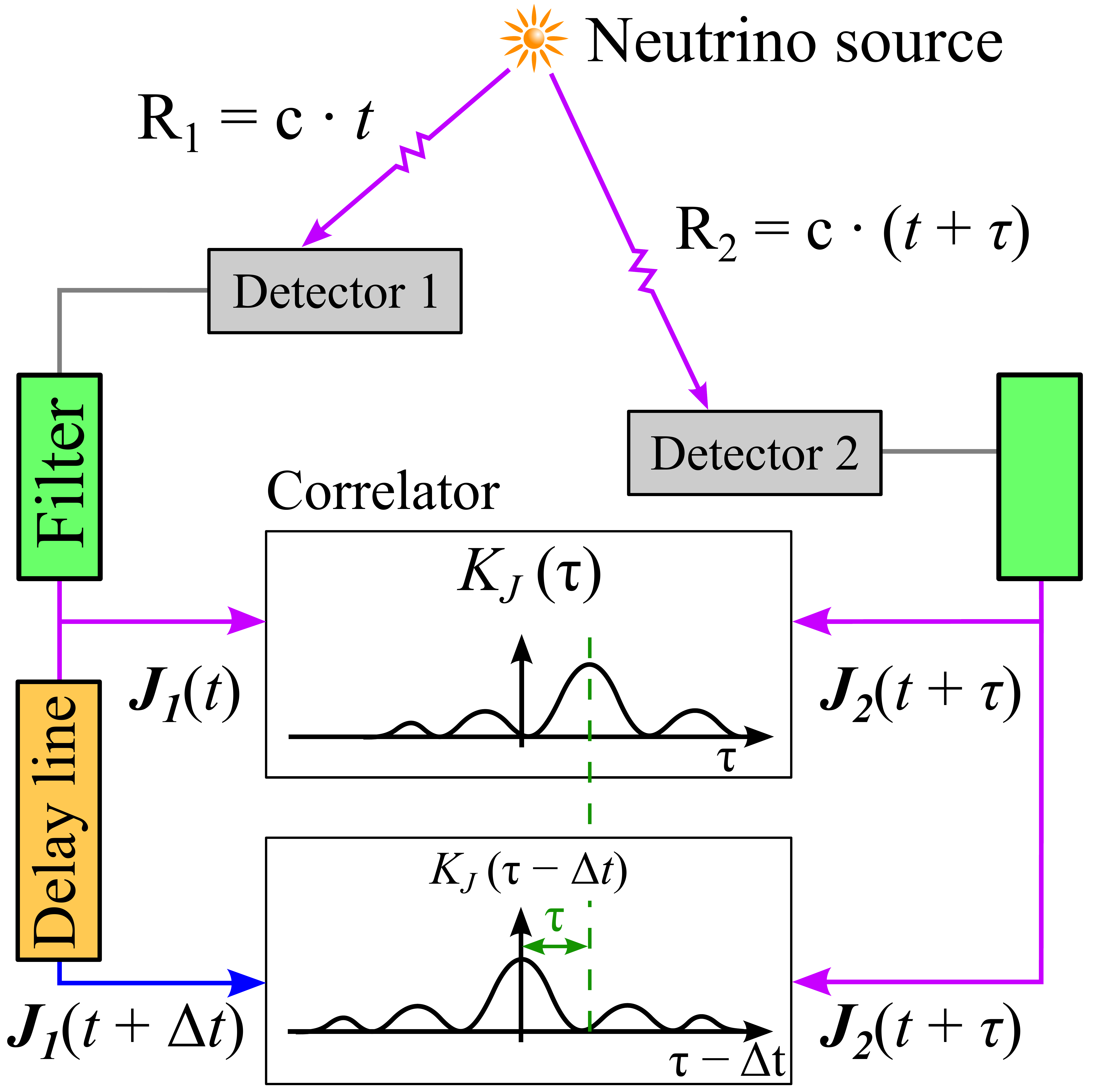}
\end{center}
\caption{Schematics of neutrino source passive location. See explanation in the text.}
\label{fig1}
\end{figure}

Let us consider the idea of interference location \cite{bib-7,bib-8} based on the simple scheme of spaced-apart neutrino detectors (Fig.~\ref{fig1}). Note that by "neutrino detector" we mean a system, in which a number of measurable nuclear physical transformations, causally related to the detected neutrino, takes place.

Within the framework of such statement of the problem we are interested in an interference correlation of neutrino intensities generated by the wave functions $\psi_1$ and $\psi_2$, which describe the neutrino motion from a source to the detector~1 and detector~2, respectively (Fig.~\ref{fig1}). In our case, a radially divergent wave from a source at the distance R, which significantly exceeds the source size, is described by the normalized spherical wave function. If the emission process lasts long enough, this function has the form of monochromatic spherical wave

\begin{equation}
\psi \left( \vec{R},t \right) \equiv  \frac{f \left( \theta \right)}{ \sqrt{4 \pi} R} \exp \left[ i \left(  { \vec{p}} { \vec  {r}} -  Et \right) / \hbar + \varphi \right]
\label{eq1}
\end{equation}

\noindent where $f(\theta)$ is the angular distribution amplitude of the radiation from source which depends on the angle $\theta$ of divergent spherical wave; $E$ and $\vec{p}$ is the neutrino energy and momentum; $\varphi$ is the initial phase of wave function; $R$ is the distance between neutrino source and detector.

Since the time of each neutrino production is sufficiently short, the wave function differs from a monochromatic wave and looks like

\begin{equation}
\psi  \left( { \vec{R}},t \right) = f \left( \theta \right) \frac{ F \left( \vec {R},t \right)}{ \sqrt{4\pi} R}  e^{i \varphi}
\label{eq2}
\end{equation}

Then the pulse sequence of neutrinos coming one by one to the detector~1 corresponds to the following total wave function

\begin{align}
\xi_{1} \left( t \right) &= \sum _{n}  {\psi_{1} \left( R_{1},t - t_{1n} \right) } = \nonumber \\
& \qquad {}= \sum _{n} { {  \frac{f \left( \theta_{1} \right) }{ \sqrt {4 \pi} R_{1} } F \left( R_{1},t - t_{1n} \right) } e^{i \varphi_{1}} }
\label{eq3}
\end{align}

\noindent and the signal intensity measured by the detector~1 is determined by the correlation

\begin{equation}
J_{1} \left( t \right) = \varepsilon_{1}\left \vert \xi_{1} \left( t \right) \right \vert ^2
\label{eq4}
\end{equation}

\noindent where $\left \vert \xi_1 (t) \right \vert ^2$ is the neutrino flux intensity in the detector~1; $\varepsilon_1$ is the registration efficiency of neutrino events by the detector~1. 

By analogy, the neutrino pulse sequence in the detector~2 corresponds to the following total wave function

\begin{align}
\xi_{2}  \left( t+ \tau \right) & = \sum _{n} {\psi_{2} \left( R_{2}, t - t_{2n} + \tau \right) } = \nonumber \\
{} &= \sum _{n}  { {  \frac{f \left( \theta_{2} \right) }{ \sqrt {4 \pi} R_{2}} } F \left( R_{2}, t - t_{2n} + \tau  \right) } e^{i \varphi_{2}}
\label{eq5}
\end{align}

\noindent where $\tau$ is the additional delay of neutrino on the way from a source to the detector~2 with respect to the neutrino traveling time from a source to the detector~1. The signal intensity measured by the detector~2 is determined by expression

\begin{equation}
J_{2} \left( t+\tau \right) = \varepsilon_{2} \left \vert \xi_{2}  \left( t + \tau \right) \right \vert ^{2}
\label{eq6}
\end{equation}

At first let us consider the operation of correlation scheme disregarding the influence of filters 1 and 2 and delay line shown in Fig.~\ref{fig1}. The cross-correlation function of signal intensities measured by detectors looks like

\begin{align}
K_{J}  \left( \tau \right) & = \left \langle J_{1}  \left( t \right) J_{2}  \left( t + \tau \right) \right \rangle - \left \langle J_{1}  \left( t \right) \right \rangle \left \langle J_{2}  \left( t + \tau \right) \right \rangle = \nonumber \\
{} &= \left\langle \varepsilon_1 \varepsilon_2 \right\rangle  \left \langle \xi_{1}  \left( t \right) \xi_{1}^{*}  \left( t \right) \xi_{2}  \left( t + \tau \right) \xi_{2}^{*}  \left( t + \tau \right) \right \rangle - \nonumber \\
{} & - \left \langle J_{1}  \left( t \right) \right \rangle \left \langle J_{2}  \left( t + \tau \right) \right \rangle
\label{eq7}
\end{align}

Since the sequence of randomly distributed pulses measured by each of the detectors is the Gaussian process, Eq.~(\ref{eq7}) can be transformed according to a known rule 

\begin{equation}
\left \langle x_{1} x_{2} x_{3} x_{4} \right \rangle = \sum _{i \neq j \neq k \neq l = 1} ^{4} { \left \langle x_{i} x_{j} \right \rangle  \cdot  \left\langle x_{k} x_{l} \right \rangle }
\label{eq8}
\end{equation}

As a result, the cross-correlation function for signal intensities (\ref{eq7}) measured by the two identical detectors has  the following form

\begin{align}
K_{J}  \left( \tau \right) & = \left\langle \varepsilon_{1} \right \rangle \left\langle \varepsilon_{2} \right \rangle  \left[ \left\langle \xi_{1}  \left( t \right) \xi_{2}^{*}  \left( t \right) \right\rangle \left\langle \xi_{2}  \left( t + \tau \right) \varepsilon_{2}^{*}  \left( t + \tau \right) \right \rangle \right. + \nonumber \\
{} &  + \left \langle \xi_{1}  \left( t \right) \xi_{2}  \left( t + \tau \right) \right\rangle \left\langle \xi_{1}^{*}  \left( t \right) \xi_{2}^{*}  \left( t + \tau \right) \right \rangle + \nonumber \\
{} &  + \left \langle \xi_{1}  \left( t \right) \xi_{2}^{*}  \left( t + \tau \right) \right\rangle \left. \left\langle \xi_{1}^{*}  \left( t \right) \xi_{2}  \left( t + \tau \right) \right \rangle \right] - \nonumber \\
{} & - \left \langle J_{1}  \left( t \right) \right \rangle \left \langle J_{2}  \left( t + \tau \right) \right \rangle
\label{eq9}
\end{align}

For the Gauss stationary processes the reciprocal moments $\left\langle \xi_i (t) \xi_i (t+\tau \right\langle$, $\left\langle \xi_i^{*} (t) \xi_i^{*} (t+\tau \right\langle$ of the random events $\xi_i (t)$ and $\xi_{j \neq i} (t+\tau)$ are zero. As a result, we have

\begin{align}
K_{J}  \left( \tau \right) & = \langle \varepsilon_{1} \rangle \langle \varepsilon_{2} \rangle \left \langle \xi_{1}  \left( t \right) \xi_{2}^{*} \left( t + \tau \right) \right \rangle \left \langle \xi_{1}^{*}  \left( t \right) \xi_{2}  \left( t + \tau \right) \right \rangle = \nonumber \\
{} &= \left \langle \varepsilon_{1} \right \rangle \left \langle \varepsilon_{2} \right \rangle  \left \vert \left \langle \xi_{1}  \left( t \right) \xi_{2}^{*}  \left( t + \tau \right) \right \rangle \right \vert  ^{2}
\label{eq10}
\end{align}

From~(\ref{eq10}) it follows that the intensity cross-correlation function is equal to the intensity of the amplitude cross-correlation function $\xi (t)$. For the considered stationary processes, representing the random pulse sequences in the formulas (\ref{eq4}) and (\ref{eq6}), the central moment $\left\langle \xi_1 (t) \xi_2^{*} (t+\tau) \right\langle$ is calculated in the standard way 

\begin{align}
& \left \langle \xi_{1} \left( t \right) \xi_{2} ^{*} \left( t + \tau \right) \right \rangle = \left \langle N_{\nu} \right \rangle \int \limits \limits _{ -\infty } ^{ \infty}  {\psi_{1} \left( R_{1}, t \right) \psi_{2}^{*} \left( R_{2}, t + \tau \right)  dt} = \nonumber \\
{}& = \frac {\left \langle N_{\nu} \right \rangle \left \langle f \left( \theta_{1} \right) \right \rangle \left \langle f \left( \theta_{2} \right) \right \rangle }{4 \pi R_{1} R_{2}}  \int \limits \limits _{- \infty} ^{\infty}  {F \left( R_{1} , t \right) F^{*}  \left( R_{2}, t + \tau \right) dt}
\label{eq11}
\end{align}

\noindent where $\left\langle N_{\nu} \right\rangle$ is the average number of neutrinos from a source during one measuring. 

Since the average number of neutrino events measured by any of the detectors per unit time is

\begin{equation}
\left \langle n_{i} \right\rangle = \left\langle N_{\nu} \right \rangle \left\langle \varepsilon_{i} \right \rangle \frac{\left \langle f \left( \theta_{i} \right) \right \rangle ^{2}}{4 \pi R_{i}^{2}}
\label{eq12}
\end{equation}

\noindent the correlation function of neutrino signal intensity (\ref{eq9}) will have the following dimensionless form:

\begin{equation}
K_{J} \left( \tau \right) = \left\langle n_{1} \right\rangle \left\langle n_{2} \right\rangle \left \vert \int \limits \limits _{-\infty}^{\infty}  {F \left( R_{1} , t \right) F^{*} \left( R_2, t + \tau \right) dt} \right \vert ^2
\label{eq13}
\end{equation}

Now let us analyze the operation features of considered scheme (Fig.~\ref{fig1}) and possible ways of its optimization. We expand the random variable $F(t)$ into the Fourier integral

\begin{equation}
F \left( R, t \right) = \int \limits \limits _{ -\infty} ^{\infty}  {F \left( R,\omega \right) \exp \left( i \omega t \right) d \omega}
\label{eq14}
\end{equation}

ubstituting the expansion~(\ref{eq14}) into~(\ref{eq13}), we obtain

\begin{align}
& K_{J} \left( \tau \right) = \left \langle n_{1} \right \rangle \left \langle n_{2} \right \rangle  \times \left \vert \int \limits \limits _{-\infty} ^{\infty}  \int \limits \limits _{-\infty} ^{\infty}  \int \limits \limits _{-\infty} ^{\infty}  F \left( R_{1}, \omega_{1} \right) F^{*} \left( R_{2}, \omega_{2} \right) \times \right. \nonumber \\
& \times \left.\vphantom{ \int \limits \limits _{-\infty} ^{\infty}} \exp {\left( i \omega_{1} t \right)} \exp {\left[ - i \omega_{2} \left( t + \tau \right) \right]}  dt d \omega_{1} d \omega_{2} \right \vert ^{2}
\label{eq15}
\end{align}

Then using the integral property

\begin{equation}
\frac{1}{2 \pi}  \int \limits \limits _{ -\infty} ^{\infty}  {\exp \left[ i \left( \omega _{1}  - \omega_{2} \right) t \right] dt = \delta  \left( \omega_{1}  - \omega_{2} \right)}
\label{eq16}
\end{equation}

\noindent we transform the correlation function~(\ref{eq15}) into the form

\begin{align}
K_{J} \left( \tau \right) & = \left\langle n_{1} \right\rangle \left\langle n_{2} \right\rangle \left( 2 \pi \right)^{2} \times \nonumber \\
{} & \times \left \vert \int \limits \limits _{-\infty} ^{\infty} { \left \vert F \left( R_{1}, \omega \right) F^{*} \left( R_{2}, \omega \right) \right \vert ^{2} e^{-i \omega \tau} d \omega } \right \vert ^{2}
\label{eq17}
\end{align}

This result is, in fact, a generalization of the Wiener-Khinchin theorem \cite{bib-9}, which describes the time dependence of amplitude correlation function of random process, for the case of intensities interference (i.e., the quadratic function of amplitudes) of the same random process.

Eq.~(\ref{eq17}) corresponds to the ideal correlator, which does not distort the processed signals $F(R_1, \omega)$ and $F_2^{*} (R_2, \omega)$ from a detector. In the real experiment such distortions can be taken into account by introducing of the correlator transfer function $G(\omega)$ which yields the equation ~(\ref{eq17}) in the form

\begin{align}
& K_{J} \left( \tau \right) = \left\langle n_{1} \right\rangle \left\langle n_{2} \right\rangle \left( 2 \pi \right) ^{2} \times \nonumber \\
{} & \times  \left \vert \int \limits \limits _{-\infty} ^{\infty}  { \left \vert G \left( \omega \right) F \left( R_{1}, \omega \right) F^{*} \left( R_{2}, \omega \right) \right \vert ^{2} e^{- i \omega \tau} d \omega} \right \vert ^{2}
\label{eq18}
\end{align}

Within the framework of OPERA experiment,the spectrum width $\delta \omega \approx 2 \pi / \delta t$ of the impulse signal $F(R,t)$, formed at the nuclear particles detector output, may be considered as exceeding the electronic post-processing circuit transmission bandwidth $\Delta \omega$ substantially. In this case one may assume that the correlator transfer function corresponds to a filter with a rectangular frequency window $\Delta \omega \ll \delta \omega$ and that the spectral density of each signal $F_{1,2}(R_{1,2},\omega)$ is constant within this frequency window and equals to $F_{1,2}(R_{1,2},0)$. As a result, from~(\ref{eq18}) we obtain

\begin{align}
 K_{J} \left( \tau \right) &= \left[ 2 \sqrt {2} \pi \left \langle n_{1} \right \rangle ^{{1} / {2}} \left \langle n_{2} \right \rangle ^{{1} / {2}} \left \vert F \left( R_{1}, 0 \right) F^{*} \left( R_{2}, 0 \right) \right \vert ^{2} \times \vphantom{\frac{\sin \Delta \omega}{\tau}} \right. \nonumber \\
{} & \left. \times  { \frac{\sin \left( \Delta \omega \tau \right) }{\tau} }  \right] ^{2} = \Phi ^{2} \left[ {  \frac{\sin \left( \Delta \omega \tau \right) }{\Delta \omega \tau} }  \right] ^{2}
\label{eq19}\end{align}

To enhance the usability of this formula and precision of measurements we supplement the signal processing circuit with two specialized controlled frequency filters with the guaranteed rectangular pass band $\Delta \omega_1 < \Delta \omega$ (Fig.~\ref{fig1}). Adding the delay line to one of the neutrino intensity interferometer arms (Fig.~\ref{fig1}) will result in the expression for the interference correlation function (and the correlator output signal) in the form

\begin{equation}
K_{J} \left( \tau - \Delta t \right) = \Phi ^{2} \left[ {  \frac{\sin \Delta \omega _{1} \left( \tau - \Delta t \right) }{\Delta \omega _{1} \left( \tau - \Delta t \right) } } \right] ^{2}
\label{eq20}
\end{equation}

The magnitude $K_{J}(\tau - \Delta t)$ tops at the optimal pulse-delay time $\Delta t = \tau$. Thus, obtaining the maximum value of the correlator output signal $K_{J} (\tau - \Delta t)$, we experimentally determine the optimum pulse-delay time $\tau$. Below we consider this problem within the framework of computing experiment based on the OPERA-data.

\section{The OPERA-data and results of computing experiment}

At first, we consider the computational results of computing experiment on the statistical simulation of neutrino measurement by two detectors located at different distances from a neutrino source. It is obvious that a statistics of neutrino detection must obey the cascade statistics of the Poisson compound distributions \footnote{The Bernoulli statistics of neutrino production in the decay tube in the OPERA-experiment does not change the kind of final distribution \cite{bib-9,bib-10,bib-11,bib-12}.}, so far as it is known, that the statistics of hadron multiple production in $pp$- and $\overline{p}p$-collisions on the energy interval in c.m.s. $\sqrt{s} = 20-1800~GeV$ is well described by the known cascade distribution $p(n)$ of the Neyman type, which belongs to the class of Poisson compound distributions \cite{bib-9}. We apply this distribution without the loss of generality in the simplest (for $\left \langle n \right\rangle \gg 1$) recurrent form \cite{bib-9,bib-10,bib-11,bib-12}:

\begin{equation}
p \left( n \right) = \sum _{k=0} ^{ \infty}  { \left[ {  \frac{ \left( k \left\langle p _{\nu} \varepsilon \right\rangle  \right) ^{n} \exp \left( - k \left\langle p _{\nu} \varepsilon \right\rangle  \right)}{n!} } \cdot  {  \frac{ \left\langle \lambda \right\rangle ^{k} \exp \left(  -  \left\langle \lambda \right\rangle  \right) }{k!}}  \right]}
\label{eq21}
\end{equation}

\noindent where $\langle \lambda \rangle$ is the average number of proton-target collisions; $\langle \varepsilon \rangle$ is the average number of secondary particles (pions and kaons) per one primary collision; $p_{\nu}$ is the Bernoulli probability of neutrino production due to secondary particle decay.

The average number of detected neutrino and variance have the form 

\begin{equation}
\left\langle n \right\rangle = \left\langle \lambda \right\rangle  \left\langle p_{\nu} \varepsilon \right\rangle
\label{eq22}
\end{equation}

\begin{equation}
var \left( n \right) = \left\langle n \right\rangle  \left( 1 + \left\langle p _{\nu}\varepsilon \right\rangle  \right)
\label{eq23}
\end{equation}

\begin{figure}
\begin{center}
\includegraphics[width=\linewidth]{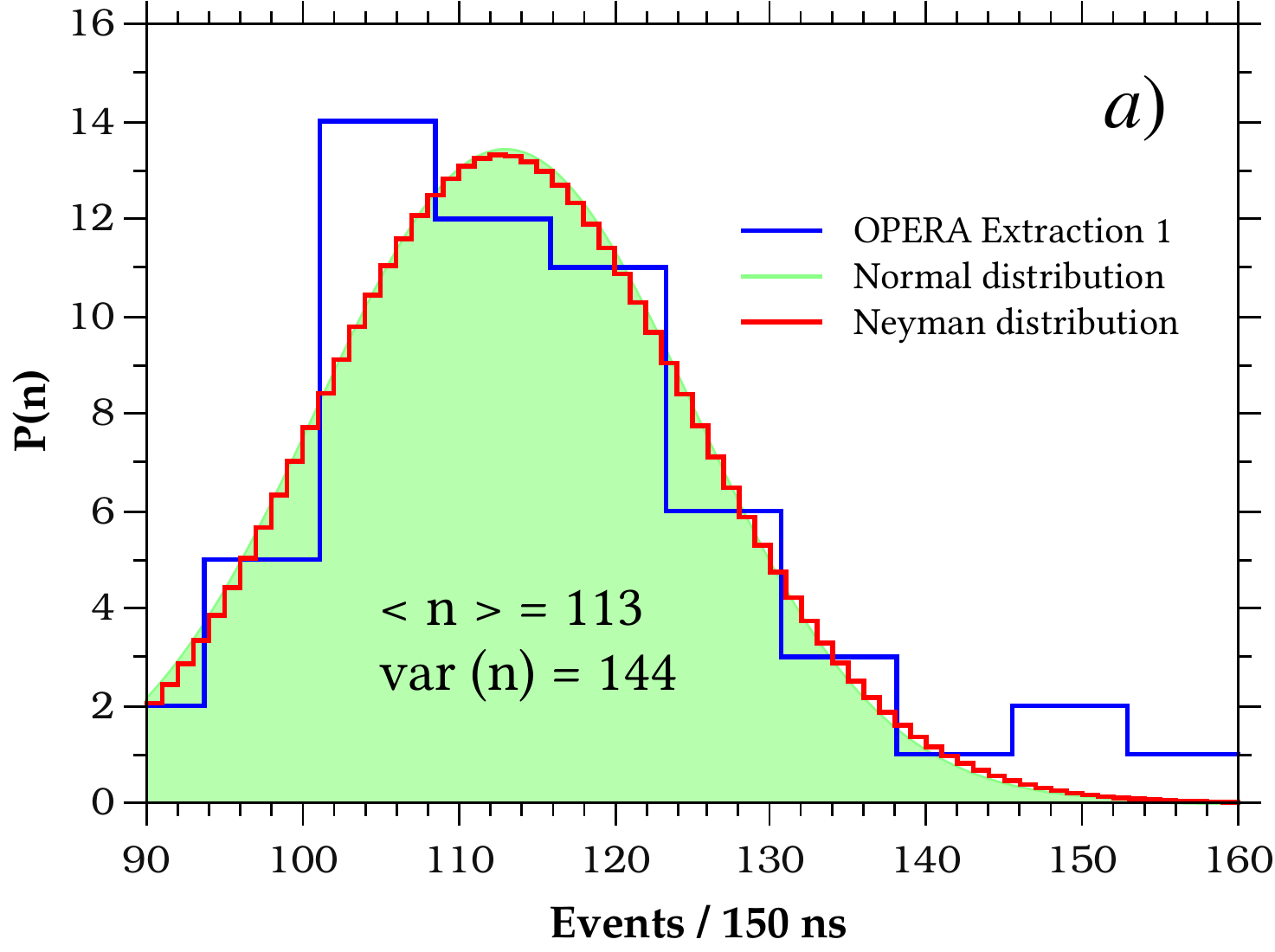}
\includegraphics[width=\linewidth]{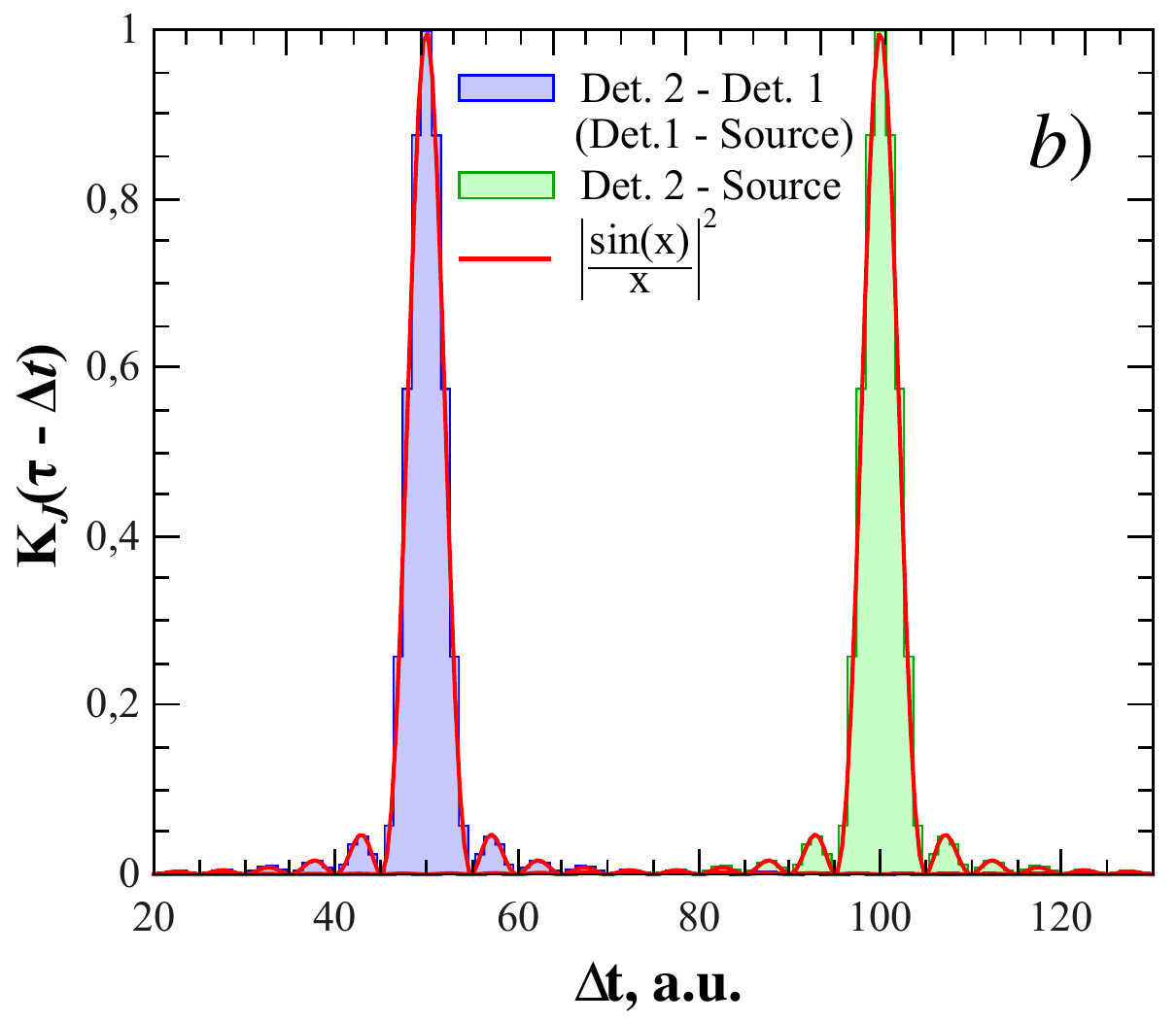}
\end{center} 
\caption{ (a) Comparison of the experimental distribution of neutrino events for the first extraction \cite{bib-1} with the average $\left\langle n \right\rangle = 113$ and variance $var (n) = 144$ (blue histogram) with the Gauss distribution (green fill area) and Neyman distribution (red histogram) with the same parameters; (b) the cross-correlation function~(\ref{eq17}) for the two samplings of neutrino events measured by two detectors spaced on the conditional time $\tau = 50 ~a.u.$ and $\tau = 100 ~a.u.$. See explanation in the text.}
\label{fig2}
\end{figure}

The experimental distribution of neutrino events obtained by the OPERA-data for the first extraction \cite{bib-1} is shown in Fig.~\ref{fig2}a. In spite of small length of neutrino event sampling, this distribution is well described by the Neyman distribution with the same parameters, i.e., $\left\langle n \right\rangle = 113$ and $var(n) = 144$.

The computing experiment on the statistical simulation of neutrino measurement by two detectors located at different distances from a neutrino source (Fig.~\ref{fig1}) was performed in the following way. Statistics of the neutrino event number is described by the Neyman distribution~(\ref{eq21}) and is set by the random number generator with the average $\left\langle n \right\rangle = 113$ and variance $var(n) = 144$ (see Fig.~\ref{fig2}). The average number of neutrinos from a source during one measurement is set so that the average number of neutrino events $\left\langle n_2 \right\rangle = 113$ (Fig.~\ref{fig2}) was registered by the detector~2 (Fig.~\ref{fig1}). We considered two locations of the detector~1 (Fig.~\ref{fig1}). In the first case the location of the detector~1 and neutrino source coincided ($\Delta t_{12} = 100 ~a.u.$), while in the second case the detector~1 was located at the distance from a source, which corresponded to the propagation time $\Delta t_{12} = 50 ~a.u.$ between the first and second detectors. The computational results for the cross-correlation function of neutrino event intensity for two variants of the first detector location are presented in Fig.~\ref{fig2}b.

It is obvious also that in the general case of two independent detectors (for example, GSL and New Detector GSL-clone), which are located at the different distances from a neutrino source, the neutrino velocity $\vartheta_{\nu}$  is determined as 

\begin{equation}
\vartheta_{\nu} = {\left \vert L_2 - L_1 \right \vert} / {\tau_{\nu}},
\label{eq24}
\end{equation}

\noindent where $\tau_{\nu}$ -- is the delay time between the neutrino signals; $L_1$ and $L_2$ are the distances between a neutrino source  and detectors, respectively.

Let us recall that in our case the value $\tau_{\nu}$ are calculated by the expression for cross-correlation function~(\ref{eq20}).

\begin{figure}
\begin{center}
\includegraphics[width=\linewidth]{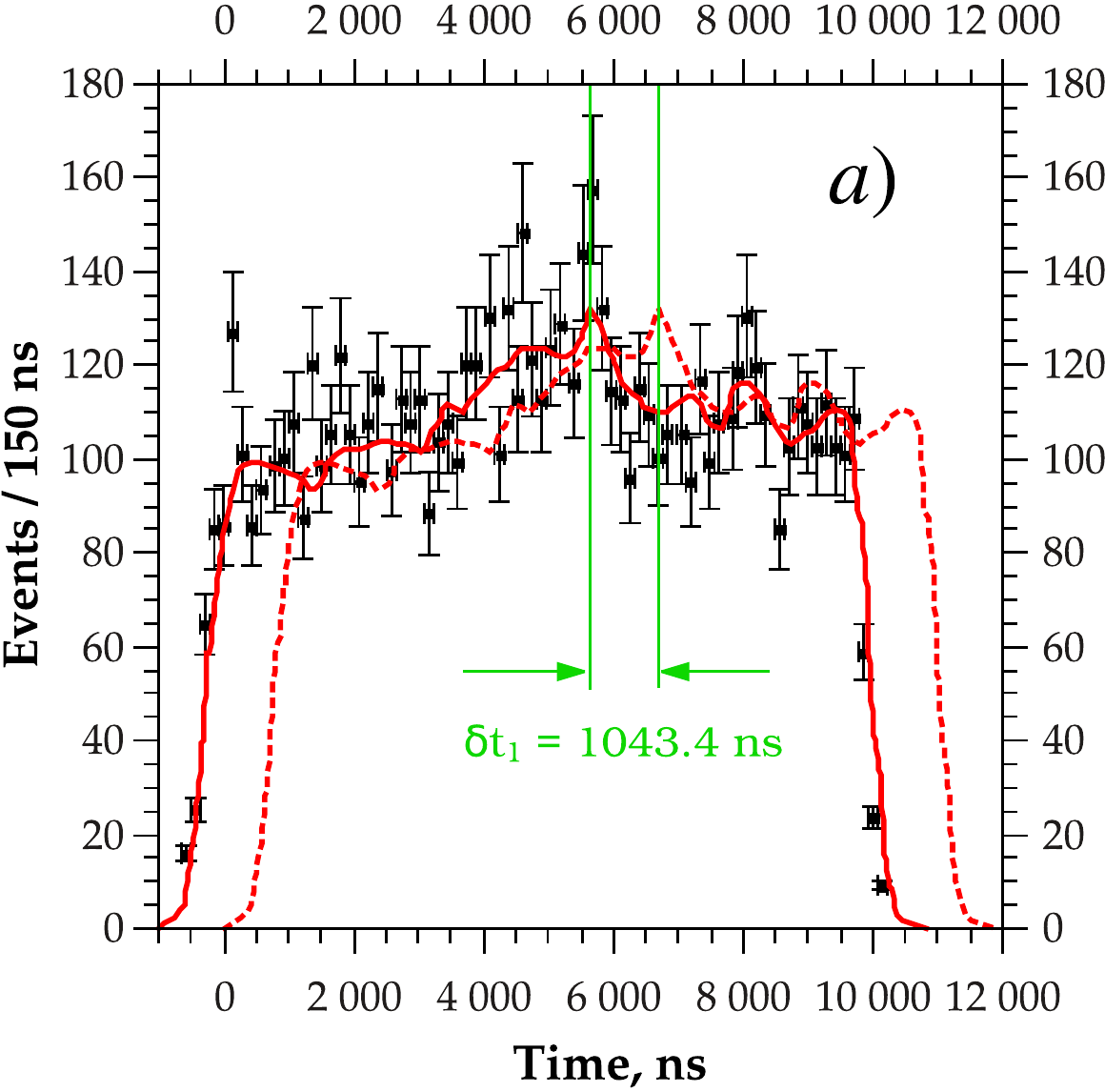}
\includegraphics[width=\linewidth]{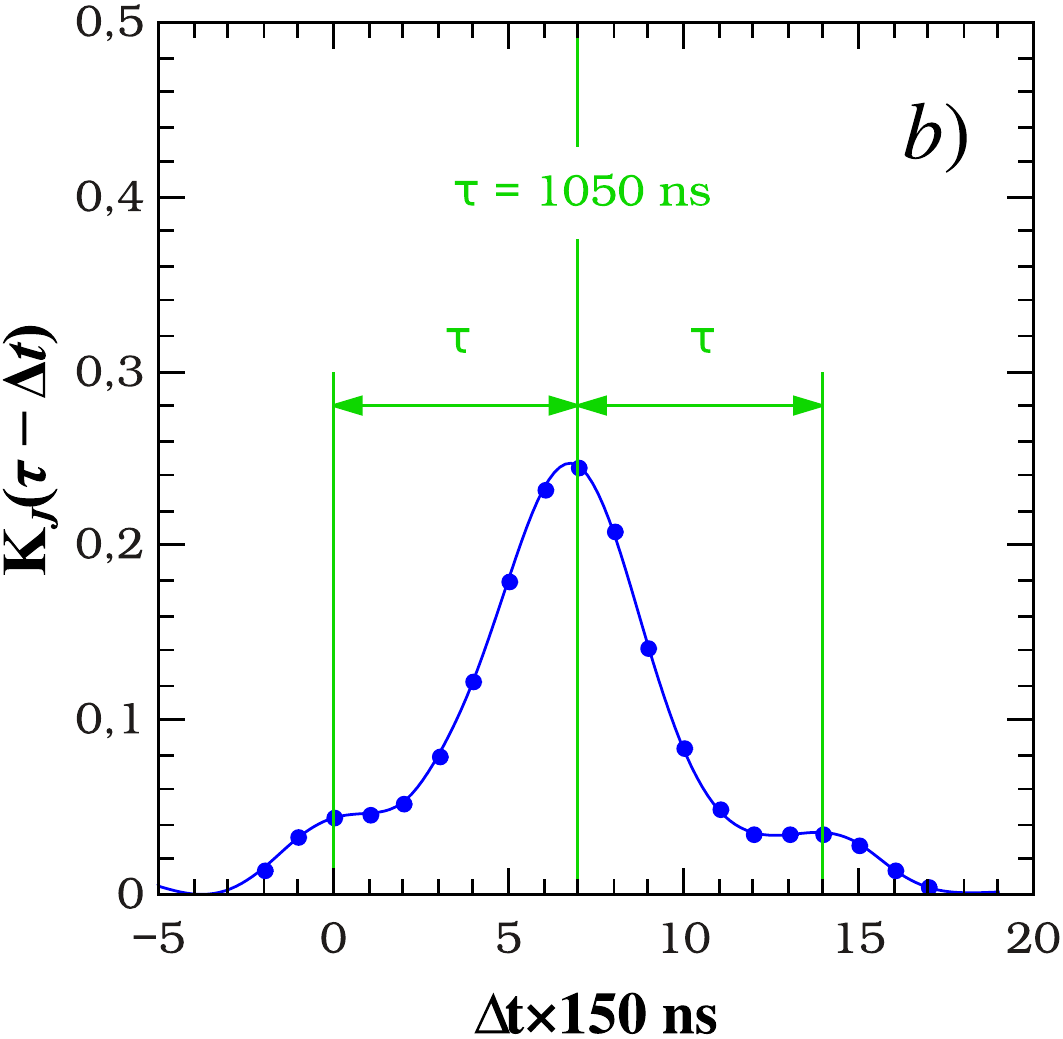}
\end{center}
\caption{(a) Comparison of the measured neutrino interaction time distribution (data points) and the proton probability density function for the first extraction before (red line) and after (dotted line) correcting for the $\delta t_1$ resulting from the maximum likelihood analysis \cite{bib-1}; (b) the cross-correlation function~(\ref{eq20}) for the two time samplings, i.e., neutrino interaction time distribution (data point in Fig.~\ref{fig3}a) and proton probability density function (dotted red line in Fig.~\ref{fig3}a) in the case of two detectors spaced on the time $\tau = 7 \times 150~ns = 1050 ~ns$. The optimum signal Fourier filtration was performed by the frequency filter with the rectangular pass band $\Delta \omega_1 = 2 \pi / \Delta t$, where $\Delta t = \tau = 7 \times 150 ~ns$.}
\label{fig3}
\end{figure}

In conclusion, we give the quantitative evaluation of cross-correlation function for the OPERA-experiment. For this purpose we suppose that the proton extraction waveform (Fig.~\ref{fig3}a) measured by a fast Beam Current Transformer (BCT) detector \cite{bib-1} and transmitted by telecommunications to GSL in the form of the proton probability density function emulates some neutrino interaction time distribution "measured" in the place of the OPERA-detector location. By virtue of the linear dependence between the proton beam intensity and average number of secondary hadrons (and, consequently, neutrinos) this supposition is sufficiently admissible. With consideration of these assumptions we will consider bellow that the signal of "the proton probability density function" travels between CERN and GSL with the speed of light, while the neutrino velocity is unknown.

Thus, based on data shown in Fig.~\ref{fig3}a we consider the problem of determining the delay time of one signal (the proton probability density function) with respect to other (the neutrino interaction time distribution), which are measured in the place of the OPERA detector location. Computational results of the cross-correlation function of neutrino events intensity for the first extraction are presented in Fig.~\ref{fig3}b. One can see that with consideration of the BCT detector resolution ($\delta t \approx 5 ~ns$, $\delta \omega \approx 1.25 \cdot 10^{9} ~Hz$) our result ($\tau = 1050 ~ns$) is in complete agreement with the OPERA result ($\delta t_1 = 1043.4 ~ns$).

\section{Conclusions} 

Based on the developed cross-correlation theory for remote location of neutrino source we offer the new arrangement of experiment, by which the precision determination of the neutrino velocity is possible. It is shown that the two-detector setup for neutrino detection makes it possible not only to determine with high accuracy the delay time between neutrino signals, but to eliminate the influence of blind analysis errors \cite{bib-1}, accompanying the procedure of performing the necessary time corrections for determination of total signal "technologically unremovable" delay time between CERN and GSL. 

It is necessary to note also a very substantial fact. Application of the cross-correlation method makes it possible to measure with high accuracy the signal time delay without using the proton pulses with steep and sharp fronts. In other words, determination of cross-correlation function does not require the signal special  preparation by the BCT-detector and allows one to use the arbitrary proton pulses with arbitrary "diffuse" fronts. This, in its turn, makes it possible to increase the statistics of neutrino events and not to reject the results of trial starts even.

\begin{acknowledgements}
One of the authors (RVD) is grateful to Dr. Vincenzo Rizi (University of L'Aquila) for the possibility to visit the Gran Sasso Underground Laboratory, where the idea of this paper arose.
\end{acknowledgements}

\bibliographystyle{apsrev4-1} 
\bibliography{SuperluminalNeutrino}

\end{document}